# Electrical Characteristics of *in situ* Mg-doped β-Ga$_2$O$_3$ Current-Blocking Layer for Vertical Devices

Sudipto Saha[1,a)], Lingyu Meng[2], A F M Anhar Uddin Bhuiyan[2], Ankit Sharma[1], Chinmoy Nath Saha[1], Hongping Zhao[2,3], and Uttam Singisetti[1, b)]

[1]*Electrical Engineering Department, University at Buffalo, Buffalo, NY, USA 14260*

[2]*Department of Electrical and Computer Engineering, The Ohio State University, Columbus, Ohio, USA 43016*

[3]*Department of Materials Science and Engineering, The Ohio State University, Columbus, Ohio, USA 43016*

The lack of p-type doping has impeded the development of vertical gallium oxide (Ga$_2$O$_3$) devices. Current blocking layers (CBL) using implanted deep acceptors has been used to demonstrate vertical devices. This paper presents the first demonstration of *in situ* Mg-doped β- Ga$_2$O$_3$ CBLs grown using metalorganic chemical vapor deposition. Device structures were designed with *in-situ* Mg doped layers with varied targeted Mg doping concentrations, which were calibrated by quantitative secondary ion mass spectroscopy (SIMS). The effectiveness of the CBL is characterized using temperature dependent current-voltage measurements using n-Mg-doped-n structures, providing crucial insight into the underlying mechanisms. To further validate the experimental results, a TCAD simulation is performed and the electrically active effective doping is found to be dependent on the Mg-doping density, offering a new perspective on the optimization of CBL performance. Breakdown measurements show a 3.4 MV/cm field strength. This study represents a significant step forward in the development of Ga$_2$O$_3$-based devices and paves the way for future advancements in this exciting field.

---

[a)]   Email: sudiptos@buffalo.edu, Tel: 701-793-7192

[b)]   Email: uttamsin@buffalo.edu, Tel: 716-645-1536

In recent years, there has been a growing trend towards developing ultra-wide-bandgap (UWBG) semiconductor materials for advanced power electronic applications as silicon-based technologies have approached their limitations [1–3]. With its high bandgap, Johnson's figure of merit (JFOM), and Baliga's figure of merit (BFOM), monoclinic β-$Ga_2O_3$ has the potential to outperform other wide bandgap devices in terms of switching efficiency and power conversion density [4–10]. These features make $Ga_2O_3$ an attractive option for high voltage and high-power power switching devices and these devices could potentially operate at elevated temperatures. Furthermore, the mature melt-growth technology used for producing high-quality, large-area $Ga_2O_3$ wafers and the ability to control n-doping over a large range sets it apart from other UWBG materials [11–20]. Various high-performance lateral $Ga_2O_3$ MOSFETs have been studied over the years with high breakdown voltages and power device figures of merit [21–26].

However, vertical devices are preferred over lateral geometries for high-voltage and high-power applications since the peak electric field is buried in the bulk to avoid surface effects [20,27]. Moreover, the breakdown voltage of vertical $Ga_2O_3$ metal-oxide-semiconductor field-effect transistors (MOSFETs) can be tuned with the drift layer thickness, i.e., a higher blocking voltage can be achieved without sacrificing chip area [20]. Owing to the lack of shallow p-type dopants for $Ga_2O_3$, the performance of vertical $Ga_2O_3$ devices is still lagging behind. The high effective hole mass and high activation energy of traditional doping species cause this p-type conductivity problem in $Ga_2O_3$ [28–33]. Researchers have used ion-implanted current blocking layers (CBL) to demonstrate vertical devices [20,34–36]. The CBL layers use deep acceptor species [36,37] to form insulating layers, which are used to engineer the electric field in vertical MOSFETs. It forms a potential barrier between the source and the drain and allows current to flow through a desired aperture in the drift layer.

Charge compensation by *in situ* epitaxy offers several advantages over the *ex situ* ion implantation process. High energy ion implantation and the subsequent thermal annealing process at high temperatures (over 1000° C for $Ga_2O_3$) results in displacement damage in the lattice, and thermal diffusion of the dopants and point defects [38–41]. *In situ* epitaxial insulating layer can be grown flexibly, avoiding the conflict of high diffusion of acceptors in $Ga_2O_3$ at high temperature and high temperature required for crystal recovery and dopant activation from *ex situ* ion implantation. Among various acceptors in $Ga_2O_3$, Mg and N impurities are usually used because of their deep acceptor nature in $Ga_2O_3$ [40,42]. Mg doping is an attractive choice because of its relatively shallow acceptor level and one of the lowest formation and activation energy compared to other cation-site acceptors from DFT calculation [43,44]. Recent EPR (Electron Paramagnetic Resonance) measurements have determined that the acceptor transition level of magnesium is located at 0.65 eV above the valence band, instead of the theoretically predicted larger values of >1 eV [45–49]. Recent experimental research also shows the semi-insulating properties of Mg acceptors in $Ga_2O_3$ as Mg can effectively capture electrons from substrate/epilayer growth interface to reduce the conductivity of n-type $Ga_2O_3$ in *in situ* Mg-doped $Ga_2O_3$ thin films [39]. However, an in-depth study of Mg incorporation in *in situ* $Ga_2O_3$ epitaxy and its electrical properties is still lacking.

In this study, we investigated the electrical properties of *in situ* Mg-doped $Ga_2O_3$ CBL for future vertical devices. Metalorganic chemical vapor deposition (MOCVD) growth was used for the designed structures. A systematic study of the electrical properties of *in situ* Mg-doped $Ga_2O_3$ CBL is investigated. We report a detailed study on the dependence of electrical behavior on the Mg doping concentration in $Ga_2O_3$ films, both at room temperature and as a function of elevated temperature. The *in situ* acceptor doping of Mg in $Ga_2O_3$ will provide versatility for designing and fabricating high-performance vertical $Ga_2O_3$ power devices.

Three $Ga_2O_3$ films were grown on commercial Sn-doped (010) oriented $Ga_2O_3$ substrates via MOCVD with three different Mg doping concentrations labeled S1, S2, and S3. The substrate surfaces were first cleaned with acetone, isopropanol, and de-ionized water prior to loading to the growth system. The UID layer and the Mg *in situ* doped layer were grown using trimethylgallium (TMGa) as the Ga precursor and high-purity $O_2$ gas for oxidation, and Argon (Ar) as the carrier gas. The Si-doped n+ layer and the Mg tail layer (UID) were grown following the previously established growth conditions using triethylgallium (TEGa) as gallium precursor [50]. Mg doping was introduced by using bis(cyclopentadienyl) magnesium ($Cp_2Mg$) as the precursor. The growth temperature was set at 880 °C for the layers using TMGa and 700 °C for the layers using TEGa, and the growth pressure was fixed at 60 Torr. The Mg doping concentration was varied by changing the Mg precursor molar flow rate (S1: 106.6 nmol/min, S2: 25.6 nmol/min and S3: 6.4 nmol/min). Under these conditions, average concentrations of $7.15 \times 10^{18}$, $8.25 \times 10^{17}$, and $2 \times 10^{17}$ $cm^{-3}$ were obtained for S1, S2, and S3, respectively (Fig. 1(b)). Quantitative secondary ion mass spectroscopy (SIMS) was utilized on a multi-layer stack, as illustrated in Fig. 1(a), to quantitatively probe the impurity profile of Mg and other impurity elements. This stack was grown on a Fe-doped substrate maintaining the same growth conditions as the S1, S2, and S3 samples. Fig. 1(b) shows the SIMS depth profiles of selected elements, including Mg, H, and C, for the sample layer stack. The Mg concentration in each sub-layer decreases monotonically as the Mg flow rate decreases. The peak Mg concentration reached in each of the three layers of Fig. 1(a) are (from top to bottom) $1.5 \times 10^{19}$, $1.75 \times 10^{18}$, and $3 \times 10^{17}$ $cm^{-3}$. At the growth temperature of 880 °C, notable Mg diffusion was observed from the SIMS depth profiles. The effect of diffusion was taken into consideration when we design the device structures in which the Mg tail layer was identified.

A schematic cross-section of the device structure for electrical testing is shown in fig. 2. The epitaxial structures for the three devices were grown on (010) β-Ga$_2$O$_3$ substrate consisted of (from bottom up) a ~0.4 μm Si-doped n$^+$ β-Ga$_2$O$_3$ layer (4.0×10$^{19}$ cm$^{-3}$), a ~0.25 μm UID layer, a ~0.25 μm *in situ* Mg doped β-Ga$_2$O$_3$ layer, a ~0.15 μm Mg tail layer followed by a ~0.5 μm Si-doped n$^+$ β-Ga$_2$O$_3$ layer (1×10$^{20}$ cm$^{-3}$). The only difference between the three fabricated samples is the Mg doping density. The average Mg concentration is 7.15×10$^{18}$ cm$^{-3}$, 8.25×10$^{17}$ cm$^{-3}$, and 2×10$^{17}$ cm$^{-3}$ for S1, S2, and S3, respectively. The fabrication started with backside etching using BCl$_3$ reactive-ion etching (RIE); a total of 1 μm thick Ga$_2$O$_3$ was etched in this step. A Ti/Au Ohmic metal stack was deposited using electron beam evaporation, followed by rapid thermal annealing (RTA) in N$_2$ at 470 $^0$C for 1 minute. The top Ti/Au/Ni Ohmic contacts were then defined using electron beam lithography, followed by RTA in N$_2$ at 470 $^0$C for 1 minute. A BCl$_3$/Ar self-aligned RIE was done to etch ~1.2 μm Ga$_2$O$_3$ to reach the bottom n$^+$ layer, with top Ni serving as the etch mask. Finally, bottom Ti/Au Ohmic metal stacks were deposited on the n$^+$ layer, followed by RTA in N$_2$ at 470 $^0$C for 1 minute. After device fabrication, current density-voltage (J-V) measurements were carried out with HP 4155B semiconductor parameter analyzer. The two probe measurements utilizing the top and back metal contacts, which (shown in fig. 2(a)) were performed and named vertical and pseudo-vertical measurements. Before testing the ohmic behavior was verified between the bottom contact and back contact. Both the vertical and pseudo-vertical measurements exhibit similar trends in the J-V results discussed in this article. An Auriga AU-5 high voltage pulsed current-voltage (I-V) setup with a 20 ns rise and fall period and a low duty cycle was used for the pulsed IV measurement to study the dynamic behavior of the Mg doped CBL. The pulse width was varied from 1 ms to 10 μs. Before measuring the devices, the system was calibrated at room temperature. In order to investigate the charge carrier compensation effect in the current blocking layer, standard reverse-biased capacitance-voltage (C-V) measurements were carried out using an Agilent 4294A precision impedance analyzer.

As seen in Fig. 3(a), all the current blocking layer structures show current blocking behavior. However, the current blocking capability varies linearly with the Mg-doping concentration, as seen in Fig. 3(b). With the highest Mg doping concentration ($7.15\times10^{18}$ cm$^{-3}$), S1 has excellent current blocking capability with a maximum forward current blocking voltage, $F_{bl}$, of 19 V and a maximum reverse current blocking voltage, $R_{bl}$, of 22.37 V. S3, with the lowest Mg doping concentration ($2\times10^{17}$ cm$^{-3}$) blocks only -0.83 V to +2.18 V. Therefore, it is evident from the experimental results that a higher Mg doping in the CBL results in stronger blocking, which is conducive to the preparation of power devices. Fig. 4 shows the temperature dependence of the current density (J) vs. voltage (V) characteristics for S1, S2, and S3 current blocking layer structures. The J-V curves of all three CBL samples show that the forward blocking capability shifts gradually toward the lower bias side with the increasing temperature representing an increase in the thermal contribution of electron transport. The reverse current densities also increase almost monotonically for all three CBLs with an increase in temperature. This is because electrons gain higher energies at elevated temperatures to the energy barrier, which attributes to the reduction in current blocking voltage. Therefore, when the temperature of the CBL structures increased, the blocking voltage range also decreased correspondingly. Fig. 5 shows the DC and pulsed J-V measurements for the three CBL devices, S1, S2, and S3, and negligible dispersion is observed. All the strucures were able to block current in the pulsed conditions, this rules out introduction of electrically active traps during *in-situ* doping.

The C-V profiles of the three tested CBLs are displayed in Fig. 6(a). The current blocking portions of all the devices exhibit almost flat C-V characteristics, supporting the idea that the current blocking regions of the devices with various Mg-doping concentrations are fully compensated, and the free charge concentration is very low. To observe breakdown, reverse biasing was applied to few of devices in each of the two samples. Devices in the S1 sample have destructive breakdown at 45 V to 59 V. Similarly, S2 sample devices have destructive breakdown between 28 V and 83 V. Figures 6(b) and 6(c) depict the reverse current density-voltage characteristics of the S1 and S2 CBLs, respectively. The maximum destructive breakdown is seen for S1 with a breakdown voltage ($V_{br}$) of -59 V and S2 with a $V_{br}$ = -83 V, giving average field strength of 2.4 MV/cm and 3.4 MV/cm, respectively.

ATLAS SILVACO simulation was carried out to understand the blocking characteristics. The materials parameters used in the simulation are listed in Table I. The SRH recombination model, Auger recombination model, and Lombardi model were all utilized in the simulation. The impact ionization parameters for β-$Ga_2O_3$ have been taken from a detailed first principle theoretical study [51]. First the doping concentration of the Mg-doped layer has been tuned to match the simulated current-blocking regions of the three structures with the experimental current-blocking capability, as shown in Fig. 8. The TCAD simulated S1 structure with Mg doping concentration of $1.35 \times 10^{17}$ $cm^{-3}$ matches the fabricated current blocking characteristics of S1 which has mean Mg doping concentration of $7.15 \times 10^{18}$ $cm^{-3}$ which leads us to the concept of effective doping. The effective doping is only 1.88% of the targeted doping of the Mg-doped layer. A similar trend is also found for S2 and S3. The effective doping is maximum for S3 (11.25%), and the effective doping rate decreases gradually with the increase of Mg-doping concentration (Table II). Understanding of the effective doping needs more careful studies. The drastic drop of effective doping for higher Mg doping could be due to formation of compensating donor defects [28] and incorporation of Mg to interstitial sites.

The conduction band diagrams of the three CBL samples in Fig. 7(a) show that S1 has the highest barrier height than S2 and S3 samples. Due to the highest Mg-doping concentration, the fermi level for S1 is closer to the valence band than S2 and S3, which results in a larger band bending under equilibrium when the fermi levels of the $n^+$ and Mg-doped layer match. As shown in Fig. 2, an Mg tail region of 150 nm is in between the top $n^+$ and the Mg-doped region due to high Mg-diffusivity in $Ga_2O_3$. The TCAD device structure also considers a graded Mg-doped region of 150 nm. As seen in Fig. 7(b) at zero bias condition, the band bending is more prominent at the $n^+$ and Mg-doped junction; the Mg activation is much higher in the interface as compared to the central region of the Mg-doped layer, where the bands are mostly flat. The trend is found in all three CBL structures. The fabricated S1 sample blocks -22.37 V to 19 V at room temperature with mean Mg doping concentration of $7.15 \times 10^{18}$ $cm^{-3}$ at room temperature. As we decrease the Mg concentration, the blocking range also decreases (Fig. 8), which can be explained by the fact that at zero bias, the conduction band barrier between the $n^+$ and the Mg-doped region decreases with decreasing doping density, as can be seen in fig. 6(b) where the red, blue and green curves represent the $7.15 \times 10^{18}$, $8.25 \times 10^{17}$, and $2 \times 10^{17}$ $cm^{-3}$, respectively.

In conclusion, this study represents the first successful fabrication of current blocking layers (CBLs) for vertical gallium oxide ($Ga_2O_3$) devices using *in situ* Mg doping at various concentrations. The electrical behavior of the devices was investigated, and the dependence of the current-voltage characteristics on the temperature and Mg doping concentration was also discussed. The results clearly demonstrate that Mg doping concentration plays a crucial role in determining the current-blocking range, with higher doping concentrations leading to improved current blocking. A TCAD SILVACO simulation was conducted to optimize the Mg-doping concentration to match the simulated current-voltage characteristics with the experimental ones, and the idea of effective doping was presented. With the increase of targeted Mg doping concentration, the effective doping decreases as the donor-type defects become more significant. These findings indicate that the *in situ* Mg doping method for developing the current blocking layer is a viable route to high-performance $Ga_2O_3$ vertical devices and will benefit the $Ga_2O_3$ device development effort. Further improvements can be obtained by optimizing the growth conditions for lower defect formation. This study is an important contribution to the power electronics community and will be of great interest to researchers and practitioners alike.

**Acknowledgments**

We acknowledge the support from AFOSR (Air Force Office of Scientific Research) under award FA9550-18-1-0479 (Program Manager: Ali Sayir), from NSF under award ECCS 2019749, and II-VI Foundation Block Gift Program. This work used the electron beam lithography system acquired through NSF MRI award ECCS 1919798.

**Data Availability Statement**

The data that support the findings of this study are available from the corresponding author upon reasonable request.

**Figure Captions:**

Fig. 1. (a) Schematics of Mg-doped homoepitaxy sample with multiple layers grown at different Mg flows (b) SIMS impurity profiles of the sample layer stack. The growth temperature is constant at 880 °C.

Fig. 2. Schematic cross-section of the current blocking layer structure fabricated on MOCVD (010) $Ga_2O_3$

Fig. 3. (a) Room temperature current density (J) vs. voltage (V) characteristics for S1, S2, and S3 current blocking layer structures (b) Forward and reverse current blocking capability of S1, S2, and S3 structures

Fig. 4. Current density (J) vs. voltage (V) characteristics at different temperatures showing variation in current blocking range with temperature for (a) S1 (b) S2 and (c) S3

Fig. 5. Pulsed current (I) vs. voltage (V) characteristics showing no DC-RF dispersion for (a) S1 (b) S2 and (c) S3

Fig. 6. (a) Capacitance vs. voltage measurements at room temperature (b) Reverse J-V characteristics of different devices of S1 sample, where the maximum destructive breakdown was observed at $V_{br}$ = -59 V. (c) Reverse J-V characteristics of of different devices of S2 sample, where the maximum destructive breakdown was observed at $V_{br}$ = -83 V.

Fig. 7. (a) Band diagram at zero bias (conduction band diagram at zero bias in inset), and (b) ionized acceptor density at zero bias for S1, S2, and S3

Fig. 8. Comparison of experimental current density-voltage characteristics with TCAD SILVACO

simulated characteristics for (a) S1 (b) S2 and (c) S3

**Table Captions:**

Table I. Design Parameters in SILVACO TCAD simulation

Table II. Mg doping concentration of fabricated CBLs and simulated Mg concentration to match fabricated current-voltage characteristics for S1, S2, and S3

**Figures:**

Fig.1:

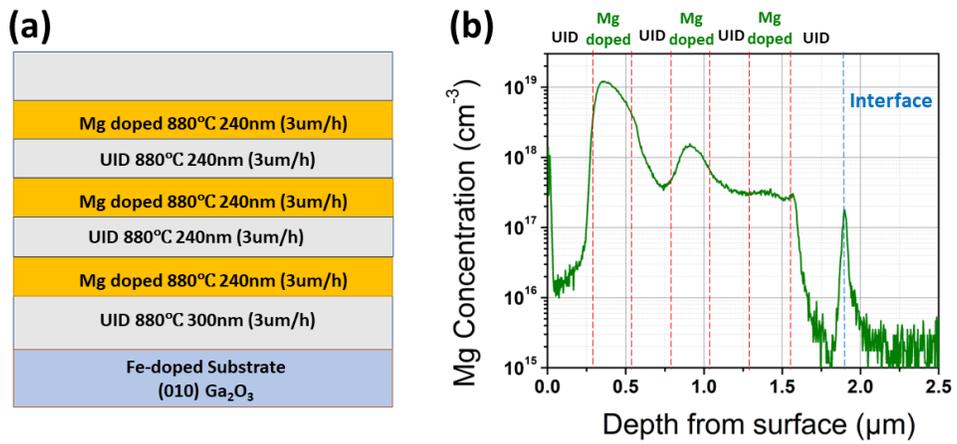

Fig.2:

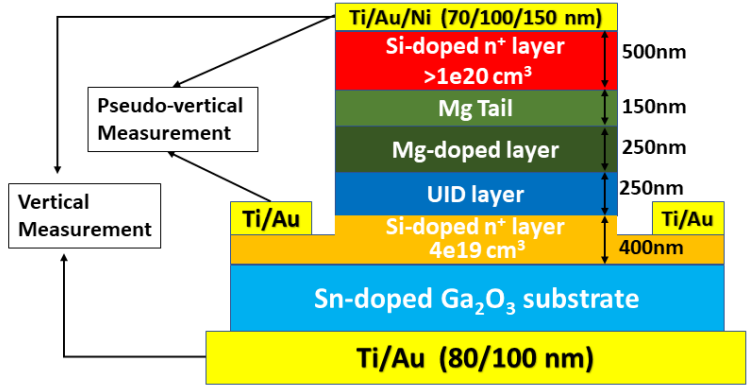

Fig. 3:

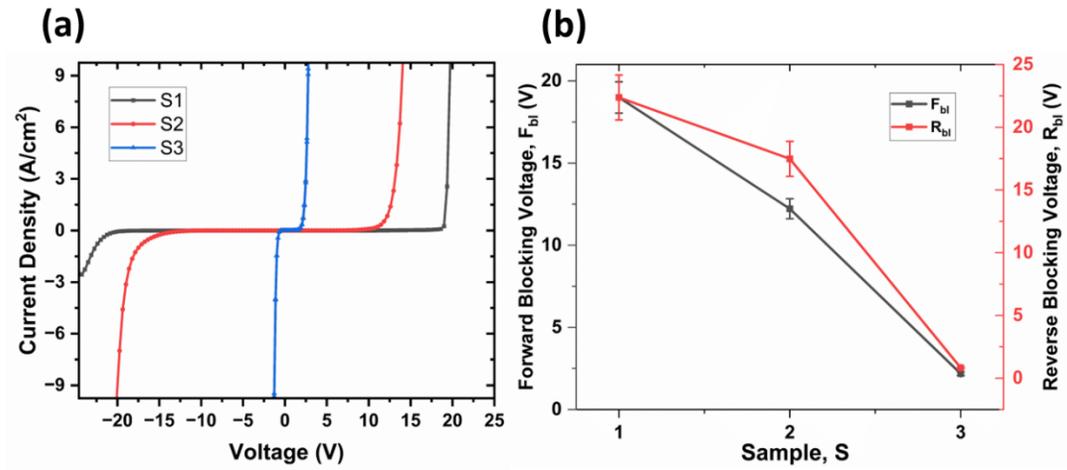

Fig. 4:

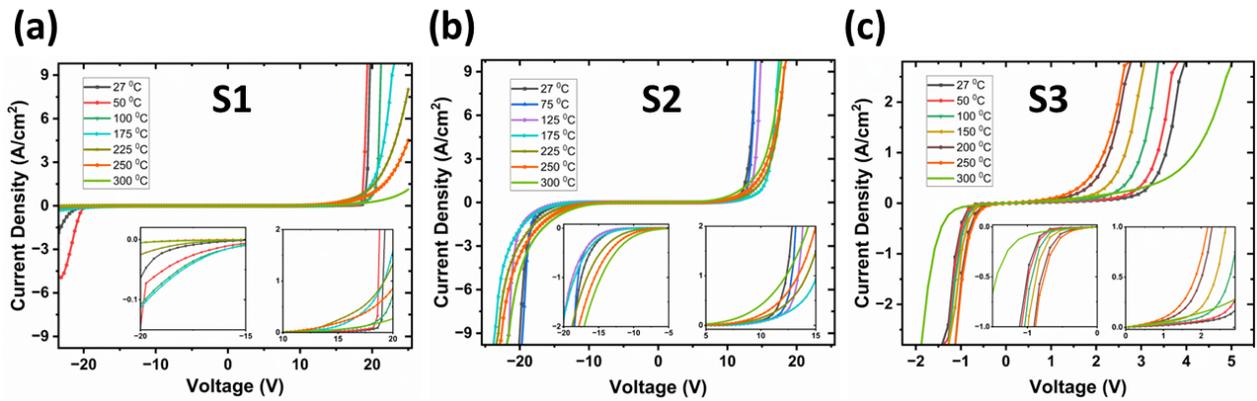

Fig. 5:

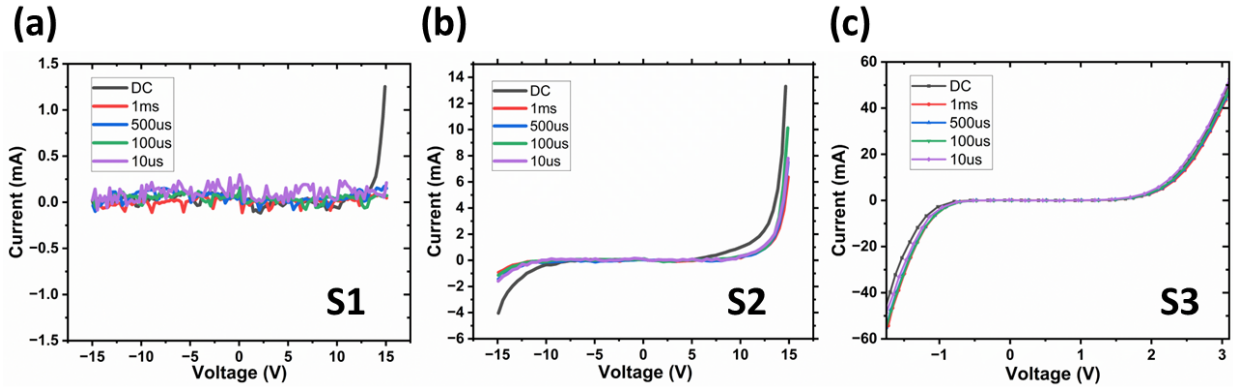

Fig. 6:

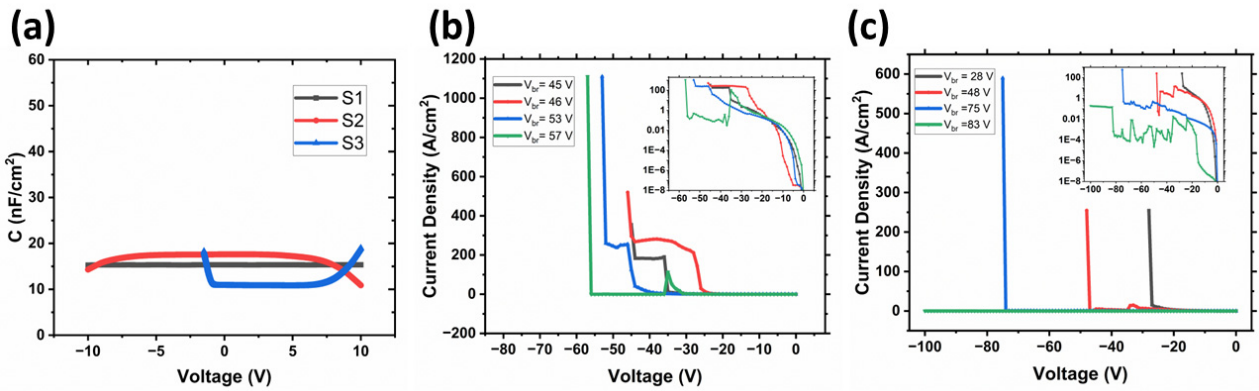

Fig. 7:

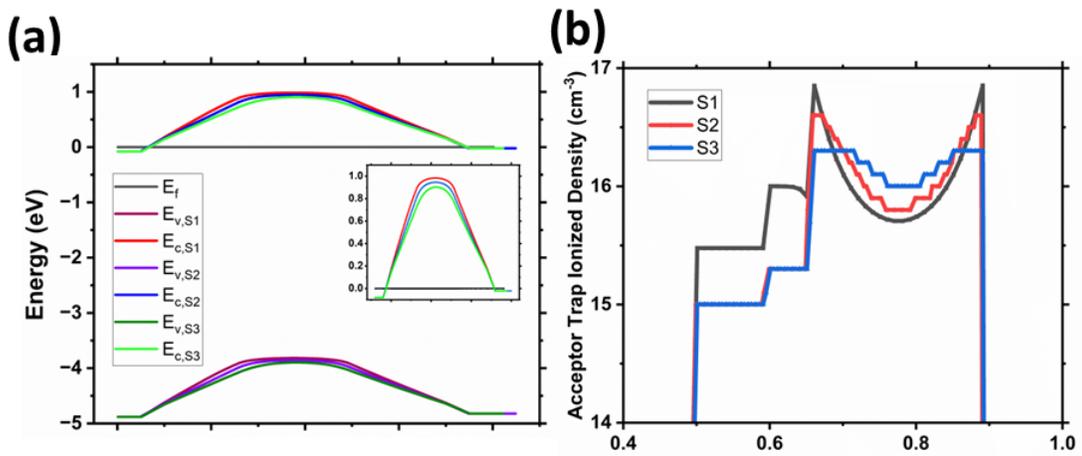

Fig. 8:

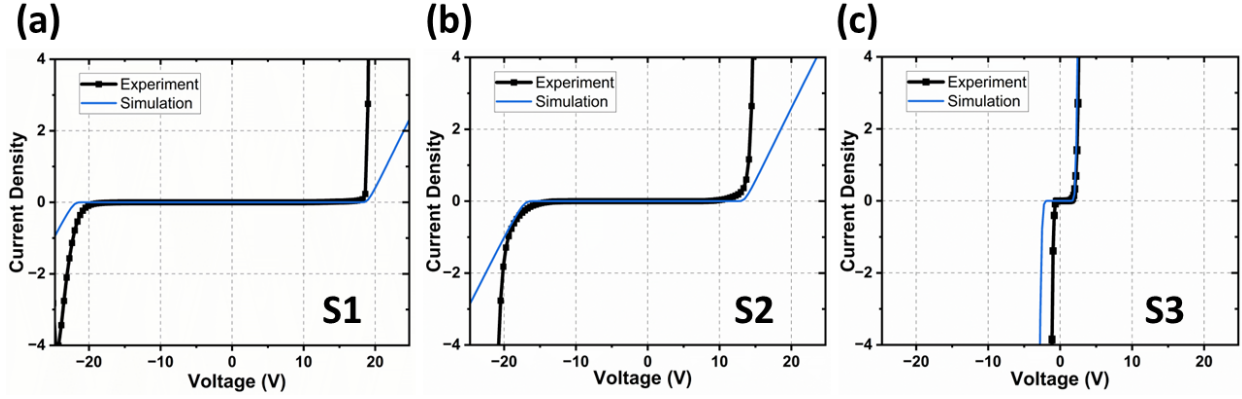

**Tables:**

Table I:

| Design Parameters | Value |
| --- | --- |
| Bandgap of $Ga_2O_3$ | 4.8 |
| Permittivity of $Ga_2O_3$ | 10.0 |
| Models used | srh, auger, cvt |
| Impact Ionization Parameters | $a=0.79\times10^6$, $b=2.92\times10^7$ |

Table II:

| Sample No | Experimental Mean Doping ($cm^{-3}$) | Simulated doping ($cm^{-3}$) | Effective doping (%) |
| --- | --- | --- | --- |
| S1 | $7.15 \times 10^{18}$ | $1.35 \times 10^{17}$ | 1.89 |
| S2 | $8.25 \times 10^{17}$ | $9.0 \times 10^{16}$ | 10.91 |
| S3 | $2.0 \times 10^{17}$ | $2.25 \times 10^{16}$ | 11.25 |